\documentclass[fleqn,twoside]{article}
\usepackage[headings]{espcrc2}

\readRCS
$Id: espcrc2.tex,v 1.2 2004/02/24 11:22:11 spepping Exp $
\usepackage{amssymb}

\usepackage{graphicx}
\usepackage[figuresright]{rotating}

\newcommand{\AmS}{{\protect\the\textfont2
  A\kern-.1667em\lower.5ex\hbox{M}\kern-.125emS}}

\title{Nuclear models on a lattice}

\author{F. de Soto\address[MCSD]{Laboratoire de Physique Subatomique et Cosmologie,
         53 avenue des Martyrs, 38026 Grenoble, France}%
        ,
        J. Carbonell\addressmark[MCSD],
        C. Roiesnel\address{Centre de Physique Th\'eorique Ecole Polytechnique, UMR7644, 91128 Palaiseau Cedex, France },
        Ph. Boucaud\address[LPTh]{Laboratoire de Physique Th\'eorique, UMR8627, Universit\'e Paris XI, 91405 Orsay-Cedex, France}       
        J.P. Leroy\addressmark[LPTh] and O. Pene\addressmark[LPTh]}
       
\runtitle{Nuclear models on a lattice}

\begin{document}


\maketitle

\section{INTRODUCTION}

We present  the first results 
of a quantum field approach to nuclear interaction models
obtained by lattice techniques.
Since Yukawa pioneer work \cite{Y_35}, these interactions are all based 
on one-boson exchange (OBE) Lagrangians.
They constitute the starting point for building the NN potentials 
\cite{NIJ_PRC_93,AV18_PRC_95,M_PRC63_01}
which, inserted in Schrodinger-like equations, 
provides an "ab-initio" description of light nuclei 
up to $A\sim10$ \cite{LightN}.

\begin{figure}[h!]
\vspace{-.5cm}
\begin{center}
\includegraphics[width=6.5cm]{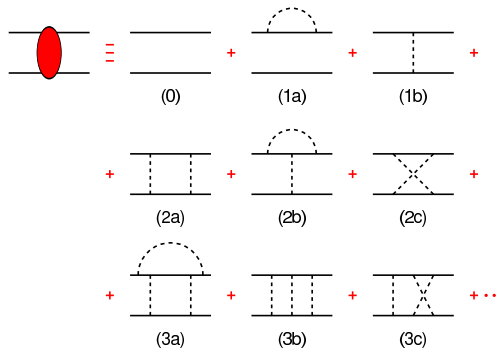}
\end{center}
\vspace{-1.cm}
\caption{Perturbative expansion of the
NN amplitude: ladder approximation corresponds to the (1b), (2a), (3b), \ldots terms.}
\label{all_diagrams}
\vspace{-0.5cm}
\end{figure}

The potential approach takes however into account only a small, though infinite, 
fraction of diagrams of the perturbative
series -- the ladder sum displayed in figure \ref{all_diagrams}.
This represents a severe restriction of the interaction,
specially taking into account the large values of the coupling  constants involved. 
Chiral inspired NN models 
\cite{W_NPB363_91,ORVK_PRC53_96,EGM_NPA671_00}, 
which can be formally distinguisehd from the OBE ones,  suffer from the same
restrictions.

Our aim is to incorporate the full content of the OBE Lagrangians
as it follows from a Quantum Field Theory (QFT) treatement of the interaction.
Lattice techniques\cite{Montvay}, 
based on a discrete Feynman path integral formulation of QFT,
provide nowadays a genuine way to solve non perturbatively such a problem. 
A similar study was undertaken in \cite{NT_PRL77_96}
in the frame of a purely scalar $\phi^2\chi$ model.

The interest of this approach is manifold. 
On one hand it  allows a comparison -- coupling by coupling --
with the results of the ladder approximation in different potential
models.
On the other hand it could provide a relativistic description of nuclear
ground states in terms of the traditional degrees of freedom -- mesons and nucleons
-- with no other restriction than those
arising from the structureless character they are assumed to have.
Of particular interest is to investigate
the possibility of obtaining the effect due to the exchange of heavy mesons
-- e.g. $\sigma,\rho$ -- in terms of interacting pions alone.

In this contribution we will focus on the renormalization effects for fermion
mass and coupling constant in case of scalar (S) and pseudoscalar (Ps) 
interaction lagrangian densities
\begin{equation}\label{Lint}
{\cal L}(x)=g_0\bar\Psi(x)\Gamma\Phi(x)\Psi(x)  
\qquad \Gamma=1,i\gamma_5  
\end{equation}
driven by the bare coupling constant $g_0$.
The scalar coupling with an additional $\lambda\phi^4$ term
-- Yukawa model --
has been investigated in the framework of the Higgs mechanism \cite{Yukawa}.

\section{THE MODEL}

The Quantum Field Theory is solved in a discrete space-time lattice of volume
$V=L^4$ and lattice spacing $a$.
All dimensional quantities are redefined in terms of
$a$  which disappears from  the formalism.
Its value can be determined only after identifying an arbitrary computed mass
to its physical value.

In terms of the dimensionless variables,
the euclidean partition functions is given by
\[  Z=\int [d\bar\psi][d\psi][d\phi] \exp\{-S_{KG}+S_F+S_{int}\}\]
where
\[ S_{KG}={1\over2}\sum_x\left\{m_0^2\phi_x^2+\sum_{\mu}(\phi_{x+\mu}-\phi_x)^2\right\}   \]
describes a real meson field with bare mass $m_0$.
The fermionic and meson-fermion coupling action is
written in the form
\[S_F+S_{int} = \sum_{xy}\bar\psi_x D_{xy}\psi_y\]
in which  
\begin{eqnarray*} 
D_{xy}&=&  (1+g\Gamma \phi_x)\delta_{xy}\cr
&-& \kappa \sum_{\mu=1}^4  (1-{\gamma}_{\mu})\delta_{x,y-\hat\mu} 
                         + (1+{\gamma}_{\mu})\delta_{x,y+\hat\mu}
\end{eqnarray*}
is the Dirac-Wilson operator.

The model depends on 3 parameters: the "hopping" parameter $\kappa$,
related to the fermion bare mass $M_0$ by 
\[ \kappa={1\over 2M_0+8}\; ,\]
the lattice coupling constant $g$ related to the interaction Lagrangian  (\ref{Lint}) by
\begin{equation}\label{g0g}
g_0 ={g\over2\kappa}
\end{equation}
and the bare meson mass $m_0$.

Our first task is to investigate
how the parameter set $(M_0,g_0,m_0)$ -- or alternatively $(\kappa,g,m_0)$ -- 
maps into the renormalized values $(M_R,g_R,m_R)$.
This task is considerably  simplified in the "quenched" approximation, 
which consists in neglecting
all virtual nucleon-antinucleon pairs originated from the meson field 
$\phi\to\bar\psi\psi$.
Because of the heaviness of the nucleon mass
this is a good approximation for the problem at hand
and has been adopted all along this work.
In this case, the meson field is trivially renormalized 
and one then has $m_R=m_0$.

\section{RENORMALIZED FERMION MASS}

In lattice calculations, the fermion propagator 
\[ G_{\alpha\beta}(x-y)=<0|\psi_{\alpha}(x)\bar\psi_{\beta}(y)|0> \]
is obtained by averaging over
all the meson field configurations -- generated by Montecarlo techniques  --
the inverse of the Dirac operator
\[ G_{\alpha\beta}(x-y) ={1\over N_{\phi}}
\sum_{\phi} D^{-1}_{\alpha x,\beta y}[\phi]\]

\begin{figure}[h!]
\vspace{-.9cm}
\begin{center}
\includegraphics[width=6.cm]{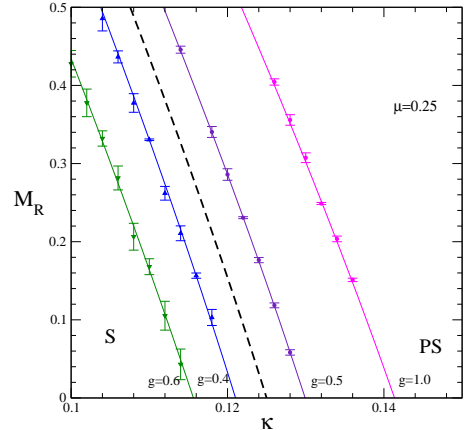}\end{center}
\vspace{-1.cm}
\caption{$M_R(\kappa)$ dependence for several values of scalar and pseudoscalar coupling.
Dashed line corresponds to the free ($g=0$) case.}\label{Mkappa}
\vspace{-.6cm}
\end{figure}

The renormalized fermion mass $M_R$ is extracted from the time-slice correlator
\[ C_{\alpha\beta}(t)=\sum_{\vec x} G_{\alpha\beta}(x)  .\]
The trajectories $M_R(\kappa)$ are displayed in figure  \ref{Mkappa}
for several values of $g$ and $m_0=0.25$.
They are monotonous functions of $\kappa$ and vanish at
the critical values $\kappa_c(g,m_0)$. 
In the region $M_R<<1$ they are  well fitted with
\begin{equation}\label{am_int}
M_R(\kappa,g,m_0)
=\frac{Z_m(g,m_0)}{2}\left[\frac{1}{\kappa}-\frac{1}{\kappa_c(g,m_0)} \right]
\end{equation}
Coefficients $Z_m(g,m_0)$ and $\kappa_c(g,m_0)$ can be calculated in perturbation theory
and provide a test of numerical calculations.
The dotted line corresponds to the free ($g=0$) case  
for which $Z_m=1$ and $\kappa_c^0\equiv\kappa_c(g=0)=1/8$.

As one can see, S and Ps trajectories lie in both sides
of the $g=0$ one with respectively $\kappa_c<\kappa_c^0$
and $\kappa_c>\kappa_c^0$. Keeping  the leading order $Z_m\approx1$
in (\ref{am_int}) one has
\[ M_R-M_0= {1\over2}\left({1\over\kappa_c^0}-{1\over\kappa_c}\right)\]
This indicates that the renormalized nucleon mass is made lighter by a scalar coupling and heavier
by a pseudoscalar one.

\begin{figure}[h!]
\vspace{-0.cm}
\includegraphics[width=7.cm]{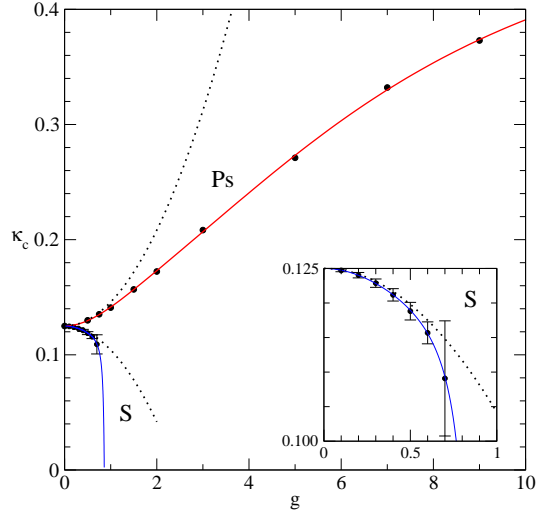}
\vspace{-.9cm}
\caption{Lattice parameter space for S and Ps couplings.
In the S case -- detailed in the zoom -- there is a critical coupling constant $g_s^c\simeq1$.}
\label{kappacg}
\vspace{-.8cm}
\end{figure}

The parameter space of physical interest is determined by the region 
$M_R(\kappa,g,m_0)>0$, i.e. the values $\kappa\in[0,\kappa_c(g,m_0)]$. 
This region is represented in figure \ref{kappacg}
for S and Ps couplings and a fixed meson mass $m_0=0.25$.
Dotted lines denote the perturbative results to $g^2$ order.
Notice that they are symmetric with respect to $k_c=1/8$.

In the scalar case the parameter space is a compact domain
limited by a critical lattice coupling constant $g_s^c\lesssim1$.
The precise determination of the $g_s^c\simeq1$ value is made difficult
by the appearence of negative eigenvalues in the Dirac-Wilson operator.
They start appearing for $g\simeq0.6$ and provoke the failure of all the algorithms 
we used in its inversion.

On the contrary the Ps coupling has a large parameter space, in
principle infinite. The spectral properties of the corresponding
Dirac-Wilson operator are very different and the negative
eigenvalues disappear for  large enough $g$ values.

Notice that $g_0$ is related
to $g$  by (\ref{g0g}) and has
sensibly larger values than those appearing in figure \ref{kappacg},
specially for the S coupling where $\kappa<\kappa_c^0=0.125$.
On the other hand these values are not yet renormalizaed  and have
no physical content. 

\section{RENORMALIZED COUPLING}

\begin{figure}[h!]
\vspace{-0.7cm}
\begin{center}\includegraphics[width=6.5cm]{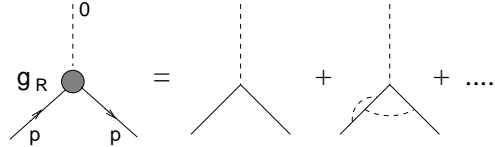}\end{center}
\vspace{-0.8cm}
\caption{Renormalized  coupling constant}\label{gR_MOM}
\vspace{-0.8cm}
\end{figure}

We have used the MOM renormalization scheme \cite{MOM} 
where the renormalized coupling constant
at a scale $\mu$ is defined as
\[ g_R(\mu)=\frac{G^{(3)}_{R,\mu}(p^2=\mu^2)}
{S_{R,\mu}(p^2=\mu^2)\; S_{R,\mu}(p^2=\mu^2)\;\Delta_{R,\mu}(0)}. \]
$S_{R,\mu}(p)$ and $\Delta_{R,\mu}(p)$ are respectively 
the fermion and meson renormalized propagators and
$G^{(3)}_{R,\mu}(p)$ the 3-point Green function represented
in Figure \ref{gR_MOM} and defined as
\[ 
G^{(3)}_{R,\mu}(p) = Z_\psi(\mu)Z^{1/2}_\phi(\mu) 
 <\psi(p) \tilde\phi(0) \bar\psi(-p) >\]
with $Z_{\psi}$ the fermion field renormalization constant.
In the quenched approximation one has $Z_{\phi}=1$

\begin{figure}[t!]
\vspace{-0.cm}
\includegraphics[width=6.5cm]{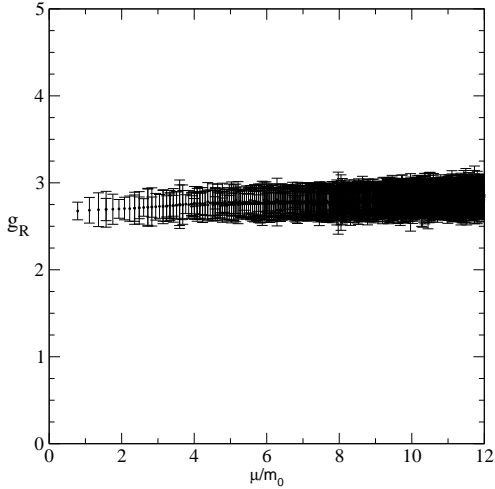}
\vspace{-1.05cm}
\caption{Momentum dependence of the scalar coupling constant obtained with $M_R/m_0=2$.}\label{g_k_S}
\vspace{-0.7cm}
\end{figure}

Our results concern the momentum dependence of the
coupling constant and the ratio $g_R/g_0$.
The first point is  illustrated in Figure \ref{g_k_S}
for the scalar coupling.  
No structure is seen up to ${\mu\over m_0}\sim12$, corresponding 
to $p\sim 6$ GeV,
although the "triviality" of the theory imposes
the existence of a Landau pole at very large momenta.
This result was found to be independent of the bare
coupling  $g_0$ and the mass ratio $M_R/m_0$.
The same behaviour has been observed with Ps coupling
for moderate $g_0$ values. 

\begin{figure}[h!]
\vspace{-0.7cm}
\includegraphics[width=6.5cm]{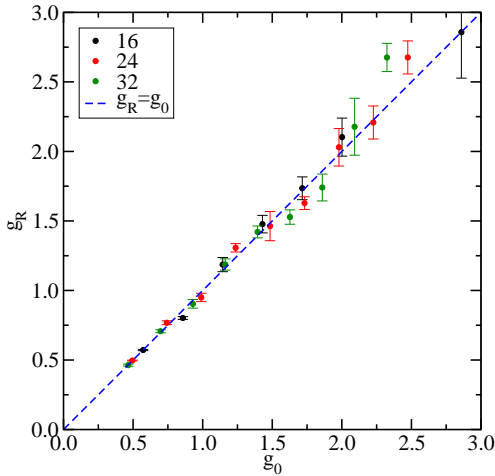}
\vspace{-1.cm}
\caption{Renormalized coupling constant versus bare values $g_0$ for the scalar case.}\label{gR_vs_g0_S}
\vspace{-0.8cm}
\end{figure}

\begin{figure}[t!]
\vspace{-0.cm}
\begin{center}
\includegraphics[width=6.5cm]{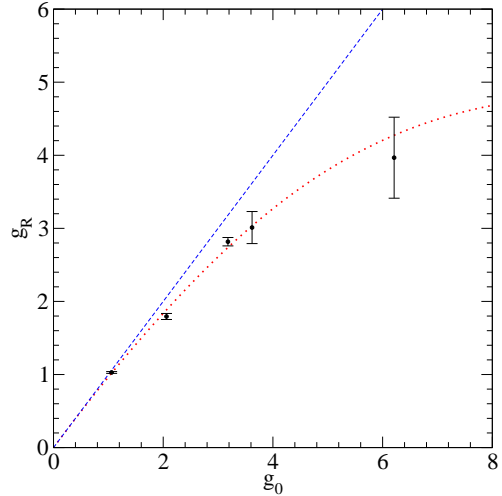}
\end{center}
\vspace{-1.2cm}
\caption{Renormalized coupling constant versus bare values $g_0$
for the pseudoscalar.}\label{gR_vs_g0_Ps}
\vspace{-.8cm}
\end{figure}

As the renormalized coupling constants exhibit very small dependence
on the momentum, we have computed  
$g_R$ as a function of $g_0$ at $p=0$. 
Results corresponding to $m_0=0.25$
are displayed in Figures \ref{gR_vs_g0_S} and \ref{gR_vs_g0_Ps}.
Here again both couplings manifest very different behaviours. 
While in the S case one has $g_R=g_0$ in all the accesible range 
($g_0={g\over2\kappa}\lesssim3$), the pseudoscalar
coupling constant $g_R$ strongly deviates from its
bare value.  
The different scales on both figures are justified 
by the effective strength of these interactions.
In the non relativistic limit, they lead to the same potential provided one has
$g_s\equiv {1\over4}\left({m_0\over M}\right)^2 g_{ps}$ .

In conclusion, we have obtained the physical parameters ($M_R,g_R$) of the 
OBE -- scalar and pseudoscalar models --
in terms of the bare quantities appearing in the Lagrangians. 
Work is in progress to obtain the dynamical properties like binding energies
of multifermion systems.

\end{document}